\documentclass[reprint, amsmath, aps, prl]{revtex4-1}
\usepackage{hyperref}
\usepackage{graphicx}
\usepackage[capitalize]{cleveref}
\begin{document}

\newcommand{\ve}[1]{\boldsymbol{#1}}

\title{Complex Crystals from Size-disperse Spheres}

\author{Praveen K. Bommineni}
\author{Nydia Roxana Varela-Rosales}
\author{Marco Klement} 
\author{Michael Engel}\email{michael.engel@fau.de}
\affiliation{Institute for Multiscale Simulation, Friedrich-Alexander-University Erlangen-N\"urnberg, Cauerstrasse 3, 91058 Erlangen, Germany}

\date{\today}

\begin{abstract}
Colloids are rarely perfectly uniform but follow a distribution of sizes, shapes, and charges.
This dispersity can be inherent (static) or develop and change over time (dynamic).
Despite a long history of research, the conditions under which  non-uniform particles crystallize and which crystal forms is still not well understood.
Here, we demonstrate that hard spheres with Gaussian radius distribution and dispersity up to 19\% always crystallize if compressed slow enough, and they do so in surprisingly complex ways.
This result is obtained by accelerating event-driven simulations with particle swap moves for static dispersity and particle resize moves for dynamic dispersity.
Above 6\% dispersity, AB$_2$ Laves, AB$_{13}$, and a region of Frank-Kasper phases  are found. The Frank-Kasper region includes a quasicrystal approximant with Pearson symbol oS276.
Our findings are relevant for ordering phenomena in soft matter and alloys.
\end{abstract}

\maketitle

\textit{Introduction.}---Dispersity~\footnote{We follow the IUPAC suggestion~\cite{Stepto2009} to replace the term `polydispersity' with the term `dispersity'. The polydispersity index, or short dispersity, is defined as the ratio of the standard deviation of the particle size distribution to the average particle size.} naturally exists in soft matter where particle geometry and chemistry can vary continuously.
It is helpful to distinguish two types of dispersities:
\emph{Static dispersity} is introduced during particle synthesis and does not change thereafter.
\emph{Dynamic dispersity} includes thermal fluctuations and the response to interactions of a particle with its environment.
Examples of dynamic dispersity are the variation of particle size through exchange of mass or charge, or the adjustment of particle shape due to forces from neighbors.

Early studies of colloids with static size dispersity predicted a terminal dispersity for crystallization between 5\% and 12\% depending on the form of the size distribution function~\cite{Dickinson1981,Barrat1986,Pusey1987,Phan1998,Auer2001,Chaudhuri2005}.
Whereas systems below terminal dispersity follow a standard phase transition into a face-centered cubic (fcc) crystal or stacking variations thereof~\footnote{We do not distinguish between fcc and its stacking variants as for example the hexagonal close-packed crystal.}, systems above were expected to fractionate into multiple coexisting fcc crystals with narrower size distribution in each crystal than the size distribution of the fluid~\cite{Bartlett1998,Sear1998,Kofke1999,Sollich2010}.
We now know that fractional crystallization does not occur in this way in experiment and simulation.
Instead, colloidal silica of dispersity 14\% coexists in the body-centered cubic (bcc) crystal, Laves phases, and the fluid~\cite{Cabane2016}.
Similarly, simulations of hard spheres with dispersity 12\% form Laves phases~\cite{Lindquist2018}, and high packing fraction and high dispersity can crystallize the AlB$_2$ structure~\cite{Coslovich2018}.
These findings were unexpected because Laves phases and AlB$_2$ are traditionally associated with binary systems~\cite{Sanders1980,Shevchenko2006,Travesset2017}.

It has been proposed that dynamic dispersity assists the formation of Frank-Kasper (FK) phases~\cite{Frank1958} and other topologically close-packed complex crystals~\cite{Iacovella2011,Lee2014,Reddy2018}.
Indeed, the FK phases A15, $\sigma$, and Laves C14 and C15 are found with micelles~\cite{Zeng2004,Cho2004,Lee2010,Huang2015,Kim2017,Baez-Cotto2018} and soft nanoparticles~\cite{Hajiw2015} where shape dispersity and size dispersity are dynamic because micelles and nanocrystal ligand shells can deform and exchange molecules.
Topologically close-packed crystals also occur in the elements Mn and U at elevated temperature~\cite{Shoemaker1978,Lawson1988,Hafner2003} where conduction electrons are mobile.

In this contribution, we investigate the crystallization of hard sphere mixtures with static and dynamic size dispersity.
While a fluid of identical hard spheres readily transitions to fcc upon densification, minor modifications of the particles strongly affects phase behavior.
Soft particles with two length scales~\cite{Dzugutov1993,Roth2000}, deformable particles~\cite{Ziherl2000,Doukas2018}, and hard particles with anisotropic shape~\cite{Damasceno2012} favor topologically close-packed or FK phases.
Interaction softening is associated with the appearance of bcc~\cite{McConnell1993,Lowen1999}.
Recent simulations of hard spheres focused on specific values of static dispersity~\cite{Lindquist2018,Coslovich2018}.
We build upon these works by applying advanced sampling techniques that allow us to study crystallization throughout the dispersity range $0\leq D\leq19\%$ and the packing fraction range $0.53\leq\phi\leq0.63$.
We determine the stability range of the Laves phase and report the first crystallization of AB$_{13}$ with hard particles in simulation.
We also discover a region of FK phases including a crystal with Pearson symbol oS276.
We finish by discussing the role of icosahedral local order and how ordering above fcc-terminal dispersity can be achieved in experiment.

\begin{figure*}
\includegraphics[width=1\linewidth]{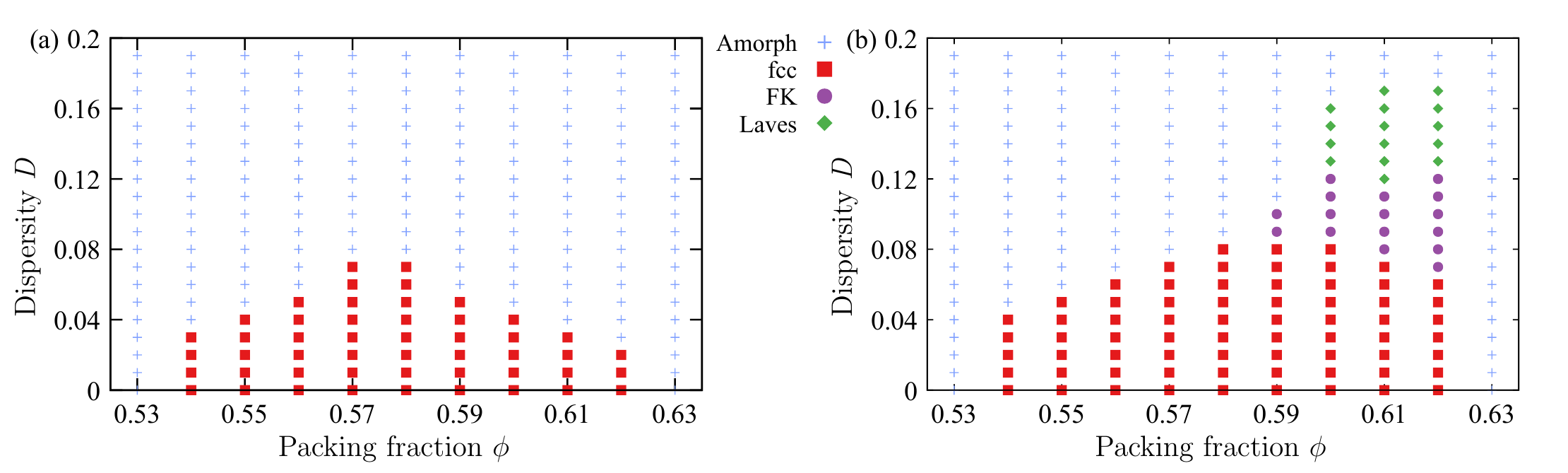}
\caption{Stability diagrams of hard spheres with static dispersity.
At each parameter pair $(\phi, D)$ the dominant phase over four simulations is marked.
The distribution $f(r)$ is used to initialize particle radii and does not change over time.
We show data for EDMD simulations (a)~without MC moves and (b)~with particle swap moves.
In each simulation, the final state at the end of the simulation is classified with the help of the radial distribution function and the bond orientational order diagram~\cite{Engel2015}.
Filled symbols represent one of the crystalline structures and pluses represent the amorphous ('Amorph', fluid or glass) state.}
\label{fig:classicandswap}
\end{figure*}

\textit{Methodology.}---We perform event-driven molecular dynamics (EDMD) simulations of hard spheres in the $NVT$ ensemble with periodic boundaries.
Spheres are initialized in a fully disordered starting configuration with Gaussian radius distribution $f(r)\propto\exp(-\frac{(2r/\sigma-1)^2}{2D^2})$, where the average diameter is $\sigma=2\langle r\rangle$.
Either a particle swap or a particle resize move may be included each time two spheres collide.
Both of these Monte Carlo (MC) moves are performed in such a way that they obey detailed balance~\footnote{See Appendix for additional text on simulation methods, transformation pathways, crystal structure identification, relative thermodynamic stability, cloud and shadow curves.}.
In a particle swap move~\cite{Kranendonk1989}, the radii of the two colliding spheres are swapped.
In a particle resize move~\cite{Zhang1999}, one radius is changed by a random number $r_1'=r_1+\Delta r$, $\Delta r\in[-s,s]$ with step size $s$.
The other radius is set to $r_2'^3=r_1^3+r_2^3-r_1'^3$, which keeps $\phi$ constant.
Resize moves sample dynamic dispersity using a semi-grand ensemble.
A move is accepted according to the Metropolis criterion with probability $\min(1,\frac{f(r_1')f(r_2')}{f(r_1)f(r_2)})$ if it does not create an overlap and rejected otherwise.
Results are obtained for $N=1000$ particles with sporadic simulations of larger systems to test for finite-size effects.
The total simulation time is $t=4\times10^5\tau$, $\tau=\sigma \sqrt{m/{k_BT}}$ with particle mass $m$, Boltzmann constant $k_B$, and temperature $T$.
We observe this hybrid EDMD-MC approach~\cite{Berthier2018} to order slightly faster than MC simulations with swap moves~\cite{Berthier2016,Ninarello2017,Brito2018}.

\textit{Static dispersity.}---We first crystallize hard spheres with static dispersity.
We compare stability diagrams for simulations with and without particle swap moves to test the effect of swaps on crystallization success.
The stability diagram without swaps has only two phases, amorphous and fcc (\cref{fig:classicandswap}(a)).
Our findings compare well with previous simulations that employ a similar simulation method~\cite{Pusey2009}.
As in that work, fcc-terminal dispersity is at $(\phi,D)=(0.58,7\%)$.
Above $\phi=0.58$, the maximal dispersity for which crystallization occurs during simulation gradually decreases.
We also tested a few selected systems in the amorphous region over the longer simulation time $t=2\times10^6\tau$ (about $10^{11}$ collisions).
But even after such long times no new crystallization event was found.
This behavior indicates a rapid slow-down of crystallization kinetics that cannot be overcome with conventional EDMD.

To access crystallization in the amorphous region, we repeat simulations in hybrid EDMD-MC by including particle swap moves at each collision.
Swap moves significantly accelerate crystallization at high dispersity and packing fraction~\cite{Berthier2016,Ninarello2017,Brito2018}.
In addition to fcc, the fluid now develops large local density inhomogeneities~\cite{Fernandez2007,Fernandez2010} and robustly and reproducibly crystallizes into Laves phases and FK phases (\cref{fig:classicandswap}(b)).
Particles do not fractionate into multiple coexisting fcc crystals according to their size but strongly mix.
The Laves phase region spans for $\phi>0.59$ up to $D=17\%$.
It includes the point $(\phi,D)=(0.595,12\%)$ where Laves phases were first seen in simulations of size-disperse hard spheres~\cite{Lindquist2018}.
The FK region is located between the Laves phase region and fcc.
We observe pronounced icosahedral local order and first-order phase transitions throughout the FK region but cannot successfully identify crystal structures.
An exception is the crystal oS276, which is discussed further below.
The speed-up from swap moves demonstrates that local rearrangements are essential to achieve crystallization in simulations of size-disperse particles.
Particles need to find appropriate locations in the unit cell that best suit their size given the overall distribution.
Above $17\%$ dispersity and 0.62 packing fraction, crystallization was once more too slow for our algorithm with swap moves (static dispersity) to access.
The metastable fluid must overcome high free energy barriers to trigger crystallization and further grow in this region.

\begin{figure*}
\includegraphics[width=1\linewidth]{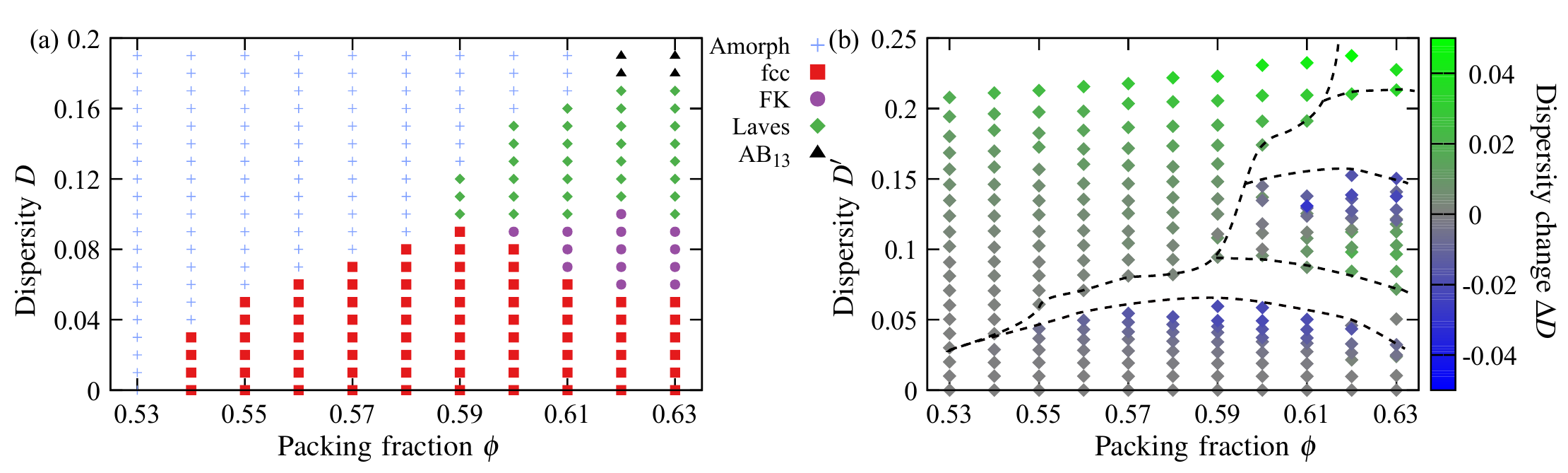}
\caption{(a)~Stability diagram of hard spheres with dynamic dispersity.
At each parameter pair $(\phi, D)$, the dominant phase over four simulations is marked.
Particle radii change over time in the vicinity of the reference distribution $f(r)$ and thus can adjust slightly to the requirements of the crystal they want to transform into.
(b)~Change in dispersity $\Delta D=D'-D$ from the value $D$ set in $f(r)$ to the value $D'$ chosen dynamically after phase transformation.
$D'$ is measured by averaging over the last frames of the simulation.
Dashed lines indicate approximate phase boundaries.
}
\label{fig:Resize}
\end{figure*}

\textit{Dynamic dispersity.}---Having established new ordering phenomena with static dispersity, we now turn to hard spheres with dynamic dispersity.
In the absence of significant interactions at low packing fraction, particle radii change due to thermal fluctuations and follow a reference distribution $f(r)$.
At higher packing fraction radii adjust to the requirements of the crystallizing system as ordering sets in.
We sample dynamic dispersity via particle resize moves at each collision and assume in our simulation algorithm that radius adjustments are subject to a free energy penalty that strives to restore the reference distribution.

The stability diagram for dynamic dispersity in \cref{fig:Resize}(a) contains ordered phases over an even larger parameter range.
AB$_{13}$ (isostructural to NaZn$_{13}$)~\cite{Bartlett1992,Eldridge1993} is a crystal structure not found in the stability diagram for static dispersity (\cref{fig:classicandswap}(b)), and the FK region is shifted to higher packing fraction.
Our simulations consistently crystallize at $\phi=0.63$, very close to random close packing, and even at $D=19\%$.
Apparently, hard spheres with dynamic dispersity crystallize much easier than hard spheres with static dispersity.

To quantify the dispersity that results from resize moves, we compute the difference between the dispersity $D$ set in the reference distribution $f(r)$ to the dispersity $D'$ chosen by the system dynamically.
The strongest shift of dispersity occurs near phase boundaries (\cref{fig:Resize}(b)).
While fluids typically retain their dispersity, crystallization into fcc lowers it.
Dispersity of systems transforming into Laves and AB$_{13}$ shifts towards the ideal values 14\% and 22\% for these crystals.
We expect similar influences of the crystal structure on the size distribution to occur in experiments that include mass or charge exchange.

\begin{figure}
\centering
\includegraphics[width=1\linewidth]{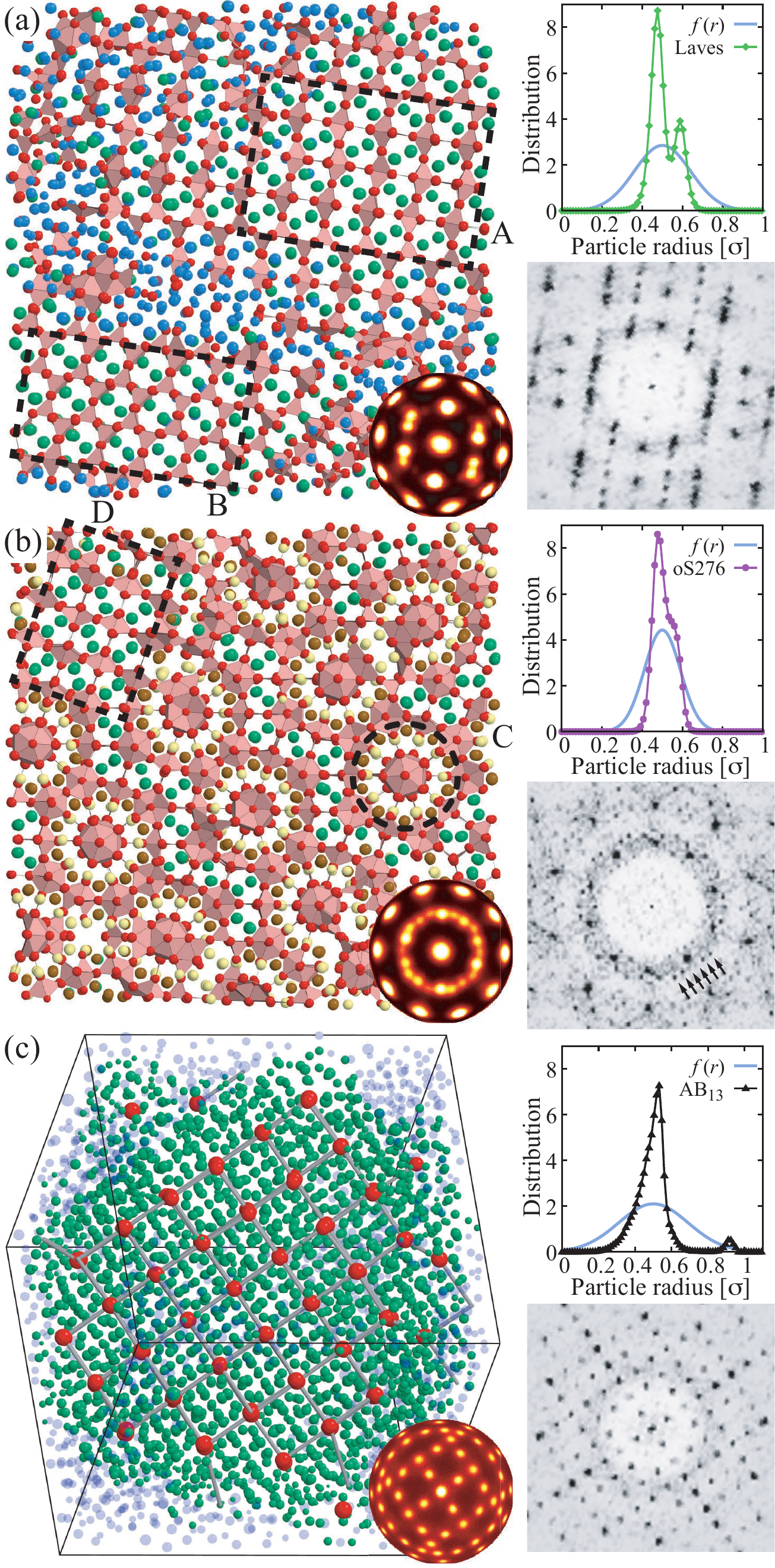}
\caption{Snapshots of (a)~Laves phase at $(\phi,D)=(0.62,12\%)$, (b) quasicrystal approximant~oS276 at $(0.61,10\%)$, and (c)~AB$_{13}$ at $(0.63,19\%)$.
Simulations are performed with resize moves.
Particles are drawn at 40\% of their size for better visibility.
In (a,b), particle colors are chosen according to coordination number $\text{CN}$: red ($\text{CN}=12$), yellow (14), brown (15), green (16), and blue ($\text{CN}\notin\{12,14,15,16\}$).
Tetrahedra of red particles form the backbone of Frank-Kasper (FK) phases.
In (c), large particles are colored red, particles near red particles green, and other particles transparent blue.
Gray lines indicate unit cells.
Bond orientational order diagrams are shown as insets.
Right side contains radius distributions (top) and diffraction images (bottom, via FFT~\cite{Engel2015}).
The reference radius distribution $f(r)$ and the radius distribution measured at the end of each simulation are compared.
}
\label{fig:Resize1}
\end{figure}

\textit{Characterization of crystal structures.}---We describe the three complex crystal structures found in our simulations in more detail.
Laves phases occupy a large area of the stability diagram in the range $D=(10\%-17\%)$ and $\phi\geq 0.59$.
In agreement with Ref.~\cite{Lindquist2018}, cubic C14 Laves and hexagonal C15 Laves coexist (\cref{fig:Resize1}(a)).
The radius distribution transforms due to resize moves into a double peak with maxima separated as expected from the binary Laves phase stability size ratio of $0.76-0.84$ and with area under the peaks following the composition AB$_2$.
Each large particle (green in \cref{fig:Resize1}(a)) is the center of a Friauf polyhedron from twelve small particles (red).
Friauf polyhedra are separated by tetrahedra (light red) that form the backbone of Laves phases and distinguish the two variants C14 (area 'A') and C15 ('B').

By comparison of bond orientational order diagrams we detect a new phase region at intermediate dispersity $D=(6\%-12\%)$ between fcc and Laves.
The symmetry of bond orientational order in simulations with $N=1000$ particles varies between icosahedral and defective decahedral, preventing us to identify crystal structures unambiguously.
We call this region the FK region because a majority of particles have coordination environments reminiscent of FK phases.
Larger simulations with up to $N=20000$ order better.
We analyze a simulation that orders particularly well as evidenced by diffraction peaks on a periodic lattice (arrows in inset of \cref{fig:Resize1}(b)).
The snapshot contains a mixture of Friauf polyhedra building blocks (green particles) and decagonal columns (area 'C'), occasionally separated by grains of Laves phase ('D').
Interpenetrating two-shell Mackay polyhedra (55 particles) form decagonal columns (64 particles) that sit at the vertices and base-center of a crystal with Pearson symbol oS276.
Only coordination numbers (CN) 12, 14, 15, 16 occur in oS276, which means it is a FK phase~\cite{Frank1958}.
The appearance of high-symmetry columns, Mackay clusters, and the mixture of building blocks from known crystals (Laves) identifies oS276 as a decagonal quasicrystal approximant~\cite{Steurer2009}.
Unfortunately, our simulations are too small to determine whether oS276 appears throughout the FK region or if there are other crystals.
In any case, we expect any crystal structure in the FK region to be much more complex than Laves phases.

Almost perfect and defect-free AB$_{13}$ crystals assemble in simulations with dynamic dispersity at high dispersity $D\geq18\%$ and high packing fraction $\phi\geq0.62$.
Large particles (red) occupy a simple cubic lattice, and small particles (green) arrange into icosahedra filling the gaps (\cref{fig:Resize1}(c)).
To mimic the 1:13 number ratio of AB$_{13}$, the radius distribution gradually self-organizes into a few large and many small particles.

\begin{figure}
\centering
{\includegraphics[width=1\linewidth]{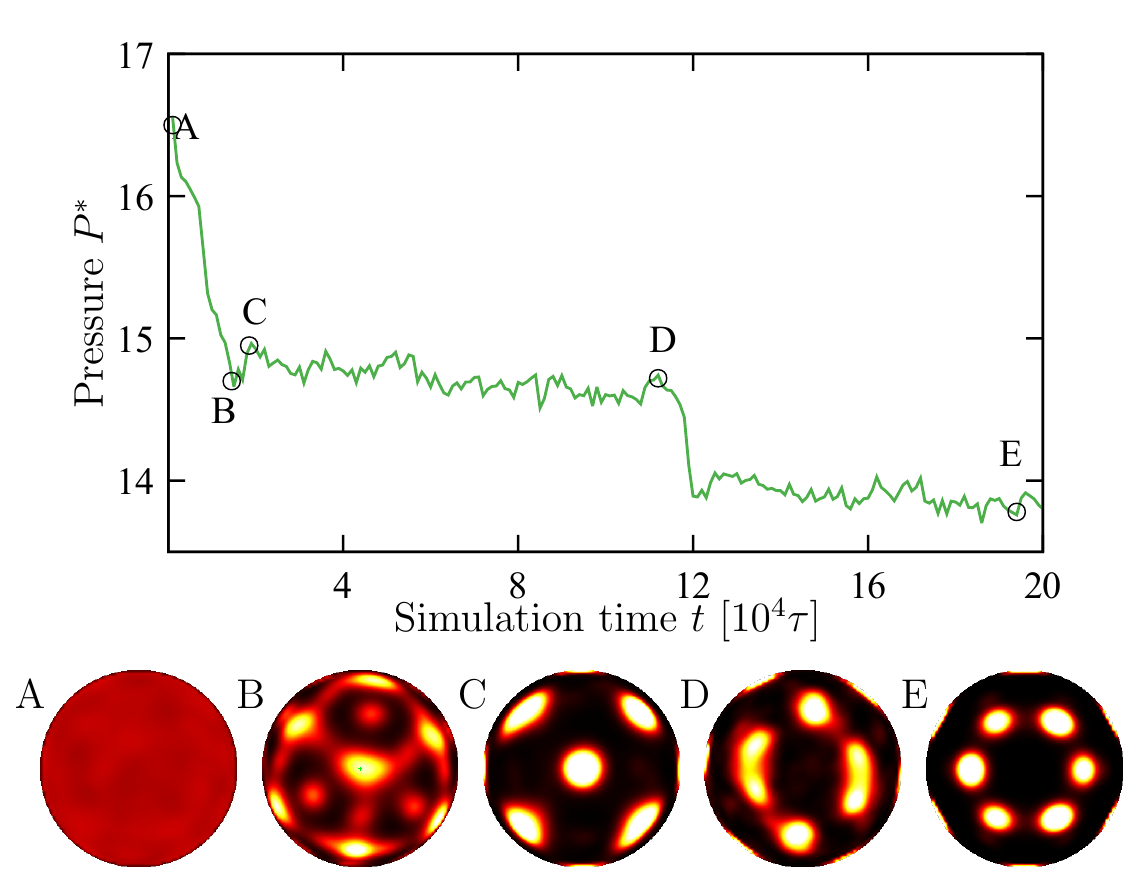}}
\caption{(Top)~Evolution of reduced pressure $P^*=P\pi\sigma^3/6k_BT$ during a particle swap simulation at $(\phi,D)=(0.60,9\%)$ that contains a transformation from fluid to fcc in multiple steps.
(Bottom)~Bond orientational order diagrams at five times during the transformation: fluid~('A'), $\gamma$-brass~('B'), bcc~('C'), defective fcc~('D'), and fcc~('E').
Additional transformation pathways can be found in the Appendix.
}
\label{fig:mstable}
\end{figure}

\textit{Transformation pathways.}---In the stability diagrams of \cref{fig:classicandswap,fig:Resize} every simulation point is mapped to a specific phase that is observed after a sufficiently long simulation time.
But not only the final simulation states are complex, also the formation into the solid frequently proceeds via multiple solid-solid phase transitions.
As metastable phases we observe bcc, and the FK phase $\gamma$-brass.
We analyze an exemplary transformation pathway in \cref{fig:mstable}.
A sharp decrease of dimensionless pressure $P^*$ marks the nucleation event (point 'A').
But the $\gamma$-brass crystal formed ('B') is short-lived.
It rapidly transforms into bcc ('C') and then to defective fcc ('D').
The defects heal, before the system finally converts into a single fcc grain ('E').
Bcc and FK phases with local icosahedral symmetry are common metastable states during crystal nucleation~\cite{Alexander1978}, but are usually not observed as easily and clearly during crystallization.
More examples of complex transformation pathways are included in the Appendix.

\textit{Discussion.}---Crystallization of mixtures is more complex and more difficult than crystallization of uniform particles.
The particles must diffuse to order successfully, which means they must overcome free-energy barriers.
Critical nucleation density increases with dispersity, and the driving force for crystallization in systems with high dispersity is particularly low.
Three strategies allow ordering systems of non-uniform nanoparticles and colloids: long time, soft interaction, and dynamic dispersity.
Natural opals made from spheres of two different sizes~\cite{Sanders1980} likely crystallize from a solution with initially continuous size distribution.
Successful opal crystallization could be the result of drying conditions that equilibrate over geological time scales much longer than typical laboratory experiments.
Soft interactions, such as flexible ligand shells~\cite{Ohara1995} and weakly decaying electrostatic forces~\cite{Cabane2016}, also assist crystallization~\cite{Fernandez2007,Fernandez2010}.
Particles with soft interactions are less strongly constrained by their neighbors and therefore diffuse more easily, speeding up crystallization.
Finally, dynamic dispersity improves crystallization because it circumvents the need for particle diffusion altogether.

Spheres have a natural tendency to develop five-fold and icosahedral local order~\cite{Frank1952,Bernal1959,deNijs2015,Wang2018}.
This tendency is enhanced by the introduction of dispersity and generally promotes glass formation~\cite{Schenk2002,Karayiannis2011,Leocmach2012,Taffs2016}.
FK phases are good candidates for crystal structures of size-disperse spheres because they combine local icosahedral order and periodicity.
Indeed, the entropic crystallization of quasi-compounds from size-disperse hard spheres, hypothesized in 1999~\cite{Kofke1999}, first observed in 2018~\cite{Lindquist2018}, and now investigated systematically in this work, mimics crystallization in alloys.
This connection relies on the observation that a continuous radius distribution $f(r)$ smears out and approximates the discrete distribution of effective atom sizes in binary and higher alloys.
The formation of diverse coordination environments is favored in both cases because it allows each particle to occupy a site that is optimally suited to its size.
Larger and longer simulations are necessary to explore the possibility of fractionation into multiple coexisting crystal phases~\cite{Sollich2010,Cabane2016} and to identify candidate unit cells in the FK region with high structural complexity.
An icosahedral or decagonal quasicrystal from size-disperse spheres derived from oS276 is a particularly intriguing prospect.

\begin{acknowledgments}
We acknowledge helpful discussions with Mahesh Mahanthappa, David Kofke, and Ludovic Berthier.
Funding by Deutsche Forschungsgemeinschaft through the Cluster of Excellence Engineering of Advanced Materials (EXC 315/2), support from the Central Institute for Scientific Computing (ZISC), the Interdisciplinary Center for Functional Particle Systems (IZ-FPS) and compute resources provided by the Erlangen Regional Computing Center (RRZE) are gratefully acknowledged.
\end{acknowledgments}

\appendix

\section{Appendix A: Simulation Methods}

\textit{Event-driven molecular dynamics.}---Hard spheres with size dispersity are simulated in event-driven molecular dynamics (EDMD) using $NVT$ simulation mode (constant particle number $N$, constant volume $V$, and constant temperature $T$).
Collision events are handled in a stable fashion~\cite{Bannerman2014} and organized in memory using a tree data structure as priority queue with O(1) complexity~\cite{Paul2007}.
Details of our EDMD implementation are described in recent work~\cite{Wang2018}.

Particles are initially placed randomly with velocities chosen according to the Maxwell-Boltzmann distribution.
The initial radii of the spheres follow a Gaussian distribution,
\begin{equation}
	f(r)\propto\exp\left(-\frac{(2r/\sigma-1)^2}{2D^2}\right),
\end{equation}
where the average diameter is $\sigma=2\langle r\rangle$, the root mean squared width of the distribution is the dispersity $D$, and all particles have equal mass $m$ independent of their radius.

While the choice of masses affects particle kinetics, it has no effect on the phase behavior of the system.
This is the case because the partition function separates, $Z(\{\ve{x}\},\{\ve{p}\})=Z(\{\ve{x}\})Z(\{\ve{p}\})$, and the positional part
\begin{equation}
	Z(\{\ve{x}\}) \propto \int\prod_{i}\prod_{j>i}H(\|\ve{x}_i-\ve{x}_j\| - (r_i+r_j)) \,d^3\ve{x}_1\cdots d^3\ve{x}_N
\end{equation}
does not depend on the choice of masses.
Here, $H$ is the Heaviside step function.
Setting the mass equal for all particles significantly simplifies momentum conservation at particle swap and particle resize moves.

Reduced pressure $P^\ast$ is computed as
\begin{equation}
	P^\ast = \frac{P \pi \sigma^3 }{6 k_B T} = \phi\left(1 + \frac{\sigma}{3} \sqrt{\frac{\pi m}{k_B T}} \frac{N_\text{pc}}{N t_\text{tot}^\ast}\right),
\end{equation}
from the number of particle collisions $N_\text{pc}$ during the simulation time window $t_\text{tot}^\ast$~\cite{Alder1960}. 

\begin{figure}
\includegraphics[width=\linewidth]{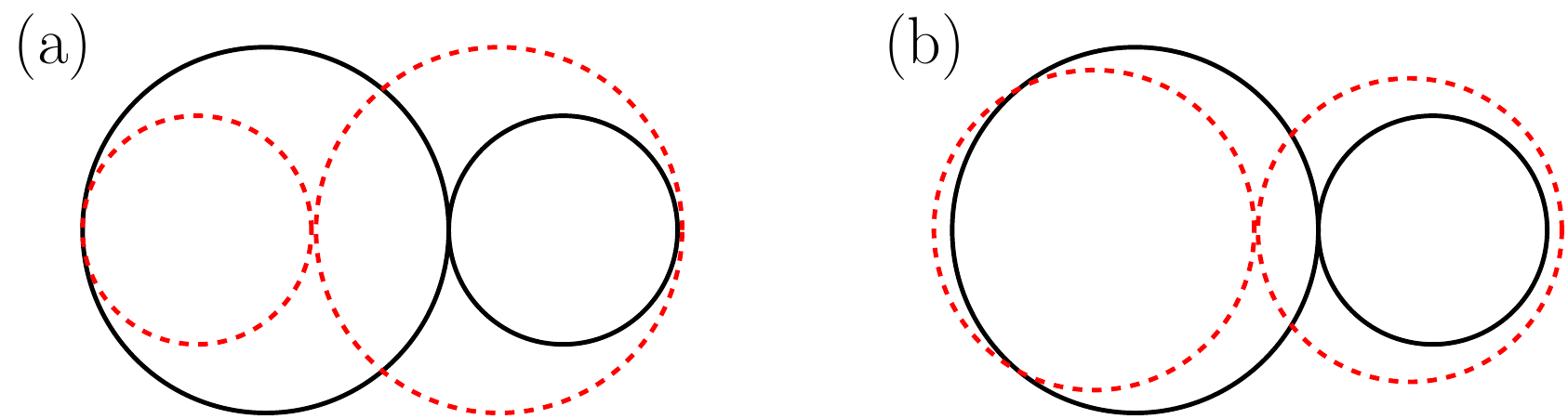}
\caption{Schematic representation of (a)~a particle swap move and (b)~a particle resize move.
In the particle swap move, the radii of two colliding spheres are swapped and their positions shifted along the connecting vector such that the segment of the connecting line covered by the two colliding spheres is kept constant.
In the particle resize move, the radius of one particle is increased while the radius of the other particle is decreased such that the sum of the volumes is kept constant.
Particles are again shifted along the connecting vector to increase the overlap of the spheres after the move with the spheres before the move.
This increase of the overlap is performed to increase the acceptance probability of the Monte Carlo move.}
\label{fig:Schematic}
\end{figure}

\textit{Particle swap move.}---Particle swapping is attempted on two colliding particles $i$ and $j$ of equal mass $m_i=m_j=m$ with radii $r_i$ and $r_j$ located at the positions $\ve{x}_i$ and $\ve{x}_j$.
After the collision, the new radii are $r_i' = r_j$ and $r_j'=r_i$ .
Furthermore, the new positions are $\ve{x}_i'=\ve{x}_i+\Delta\ve{x}$ and $\ve{x}_j'=\ve{x}_j+\Delta\ve{x}$ with shift vector $\Delta\ve{x}$ along the normalized connecting vector
\begin{equation}
	\hat{\ve{x}}_{ij}=\frac{\ve{x}_j-\ve{x}_i}{r_i+r_j}
\end{equation}
chosen as
\begin{equation}
	\Delta\ve{x} = \hat{\ve{x}}_{ij}(r_j-r_i).
\end{equation}
This choice maximizes the overlap of the spheres after the collision with the spheres before the collision (\cref{fig:Schematic}(a)).
Velocities are updated according to a regular collision of spheres with equal mass.
The new velocities are $\ve{v}_i'=\ve{v}_i+\Delta \ve{v}$ and $\ve{v}_j'=\ve{v}_j-\Delta \ve{v}$ with
\begin{equation}
	\Delta\ve{v} = \hat{\ve{x}}_{ij}(\hat{\ve{x}}_{ij}\cdot\ve{v}_{ij})
\end{equation}
and $\ve{v}_{ij}=\ve{v}_j-\ve{v}_i$.
A particle swap move is accepted if it does not generate an overlap in the system.

\textit{Particle resize move.}---Particle resizing is attempted on two colliding particles.
The new radii are set to
\begin{eqnarray}
	r_i'&=&r_i+\Delta r,\\
	r_j' &=& (r_i^3 + r_j^3 - r_i'^3)^{1/3},
\end{eqnarray}
which automatically guarantees conservation of packing fraction in the simulation box.
Here, $\Delta r$ is a random number uniformly distributed within a symmetric interval $[-s,s]$ bounded by the step size $s$.
The step size is adjusted during initialization to achieve an acceptance probability for particle resize moves of about 50\% and then kept constant.
The identity of the first particle ($i$ or $j$) is chosen randomly to ensure global balance.
Particles are shifted slightly to keep the particles in contact after the resize and increase the overlap of the spheres after the collision with the spheres before the collision (\cref{fig:Schematic}(b)).
The new positions are
\begin{eqnarray}
	\ve{x}_i' &=& \ve{x}_i + \hat{\ve{x}}_{ij} (r_i-r_i'),\\
	\ve{x}_j' &=& \ve{x}_j - \hat{\ve{x}}_{ij} (r_j-r_j').
\end{eqnarray}
Velocity updates are performed identically to velocity updates in a particle swap move.
A particle resize move is accepted if it does not generate an overlap in the system according to the probability
\begin{equation}
	P=\min\left(1,\frac{f(r_1')f(r_2')}{f(r_1)f(r_2)}\right).
\end{equation}
This equation is a Metropolis criterion, which assumes a radius distribution $f(r)$ for both particles involved in the resize move.

In other words, particle resize simulations sample the semi-grand ensemble that fixes packing fraction and the number of particles but not the size distribution.
The chemical potential $\mu(r)$ in this ensemble depends on particle radius and obeys $\exp(\mu(r)/k_B T)=f(r)$.
In the case of the Gaussian radius distribution employed in this work,
\begin{equation}
\mu(r) = -k_B T\frac{(2r/\sigma-1)^2}{2D^2}.
\end{equation}

For stability reasons of the simulation algorithm, particle resize moves that attempt to assign a negative radius or a radius larger than $5D$ are automatically discarded if they occur, which, however, is extremely rare.

\begin{figure}
\includegraphics[width=\linewidth]{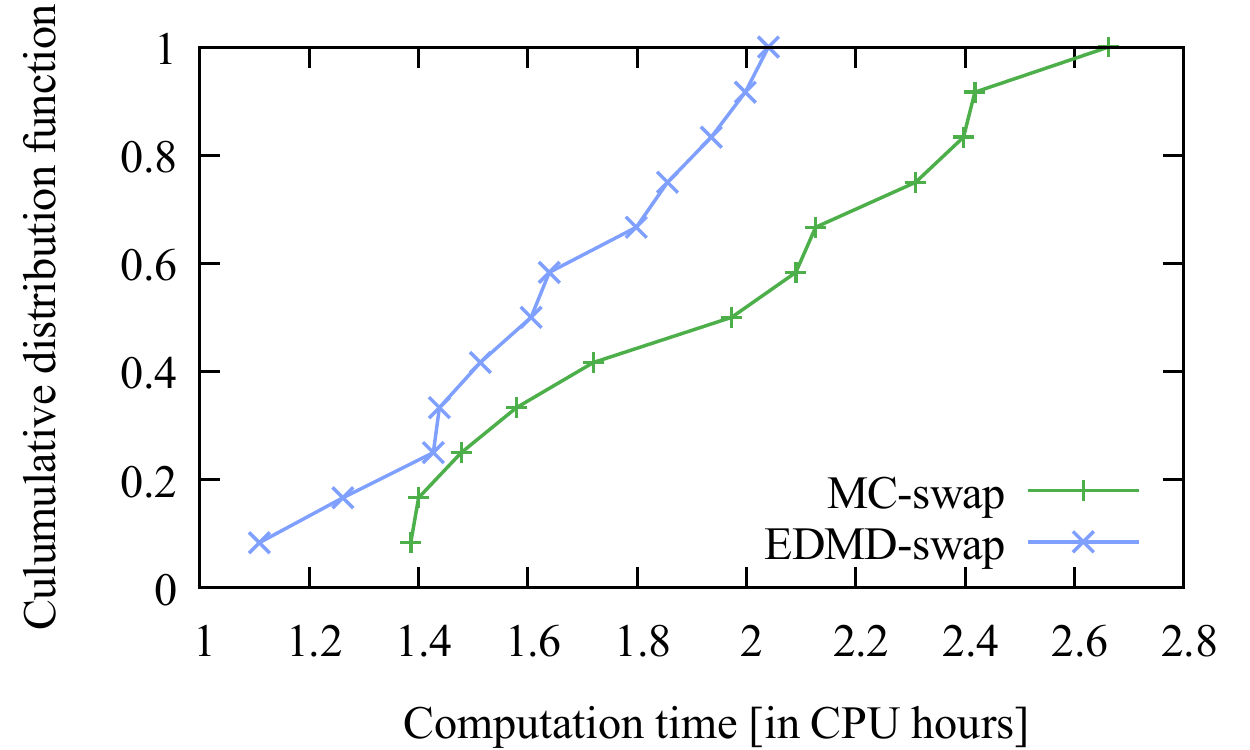}
\caption{Cumulative distribution function of computation time to crystallize a $N=1000$ system of disperse spheres at $(\phi,D)=(0.62,12\%)$ with EDMD-swap and MC-swap.}
\label{fig:Performance}
\end{figure}

\textit{Performance comparison.}---We compare the performance of the EDMD simulation method with swap moves (EDMD-swap) to conventional MC simulations with swap moves (MC-swap, \cite{Kranendonk1989,Berthier2016}).
For this comparison we measured the computation time for crystallization of a disperse, initially random system.
\cref{fig:Performance} is the cumulative distribution function of the computation time to observe nucleation from a set of twelve simulations.
Nucleation was identified from the bond orientational order diagram.
We observe a slight performance increase of 10-20\% with EDMD-swap over MC-swap.

%%%%%%%%%%%%%%%%%%%%%%%%%%%%%%%%%%%%%%%%%%%%%%%%%%%%%%%%%%%%%%%%%%%%%%%%%%%%%%%%%%%%%%%%%%%%%%%%%%%%%%%%%%%%%%%%%%%%

\begin{figure*}
\includegraphics[width=\textwidth]{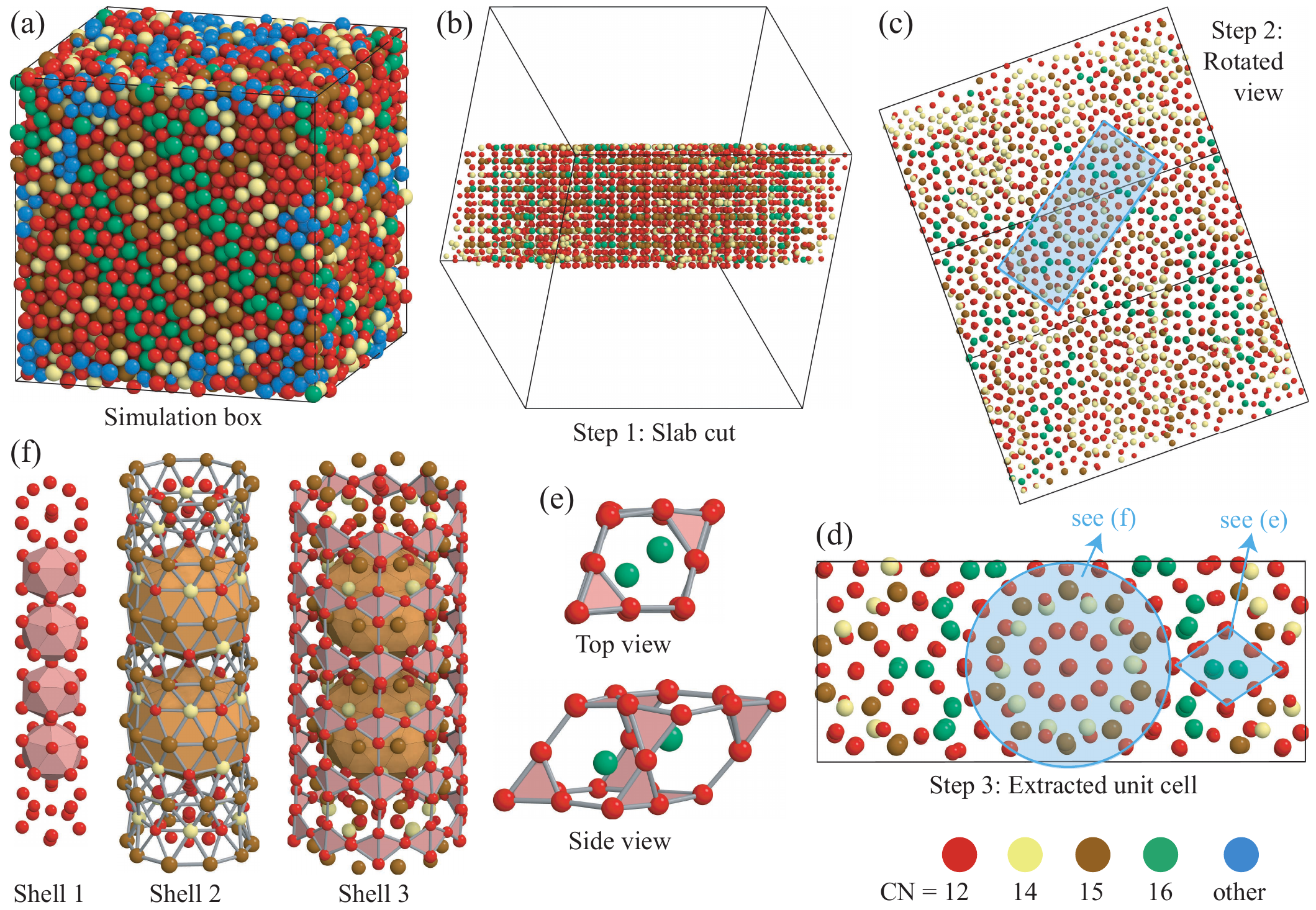}
\caption{Crystal structure identification of oS276.
The final simulation snapshot~(a) is gradually cut down towards the orthorhombic unit cell in three steps~(b-d), for details see the text.
Colors indicate the coordination number (CN) of a particle.
Two elementary building blocks are present: (e)~Laves rhombohedra and (f)~decagonal columns with embedded Mackay clusters (brown).
In (b-f), particles are drawn at 40\% of their size for better visibility.}
\label{fig:CrystalID}
\end{figure*}

\section{Appendix B: Crystal Structure Identification}

Crystal structures are identified using a combination of real space observables and reciprocal space observables as detailed in prior work~\cite{Engel2015}: radial distribution function (RDF), bond orientational order diagram (BOD), and diffraction image.
The procedure is straight-forward for the simple crystal structures fcc and bcc.
Laves phases and AB$_{13}$ are identified initially using BODs.
C14 and C15 Laves are characterized unambiguously by the coordination numbers of the particles (exclusively 12 and 16) and distinguished by the building block arrangement when viewed along a two-fold axis as shown in~\cref{fig:Resize1}(a).
AB$_{13}$ rarely has defects and the regular arrangement of the large particles (red in \cref{fig:Resize1}(c)) on a simple cubic lattice is sufficient to separate the crystal from amorphous regions.

The identification of oS276 is more complicated and proceeds in several steps:

\emph{Step 1.} Starting from the particles in the simulation box (\cref{fig:CrystalID}(a)), we extract a region where the crystal is well-formed, which is the case in the region where no blue particles are present.
We then cut out a slab perpendicular to a high-symmetry direction (\cref{fig:CrystalID}(b)).

\emph{Step 2.} The slab is rotated parallel to the high-symmetry direction.
Translational symmetry allows further extraction of a rectangular column that repeats in the structure (\cref{fig:CrystalID}(c)).

\emph{Step 3.} We analyze periodicity perpendicular to the column direction and extract a final orthorhombic box, the unit cell.
We determine directly in real space that unit cell is orthorhombic base-centered and contains 276 particles (\cref{fig:CrystalID}(d)).

Particles are colored according to their coordination number (CN).
We know from the simulation data (\cref{fig:CrystalID}(a)) that the crystal is a Frank-Kasper phase.
The fact that the extracted unit cell only has particles that obey the Frank-Kasper condition $\text{CN}\in\{12,14,15,16\}$ is a strong indication that the unit cell was extracted correctly.
A missing or redundant particle would inevitably lead to CNs that violate the Frank-Kasper condition.

\begin{figure}
\includegraphics[width=0.86\linewidth]{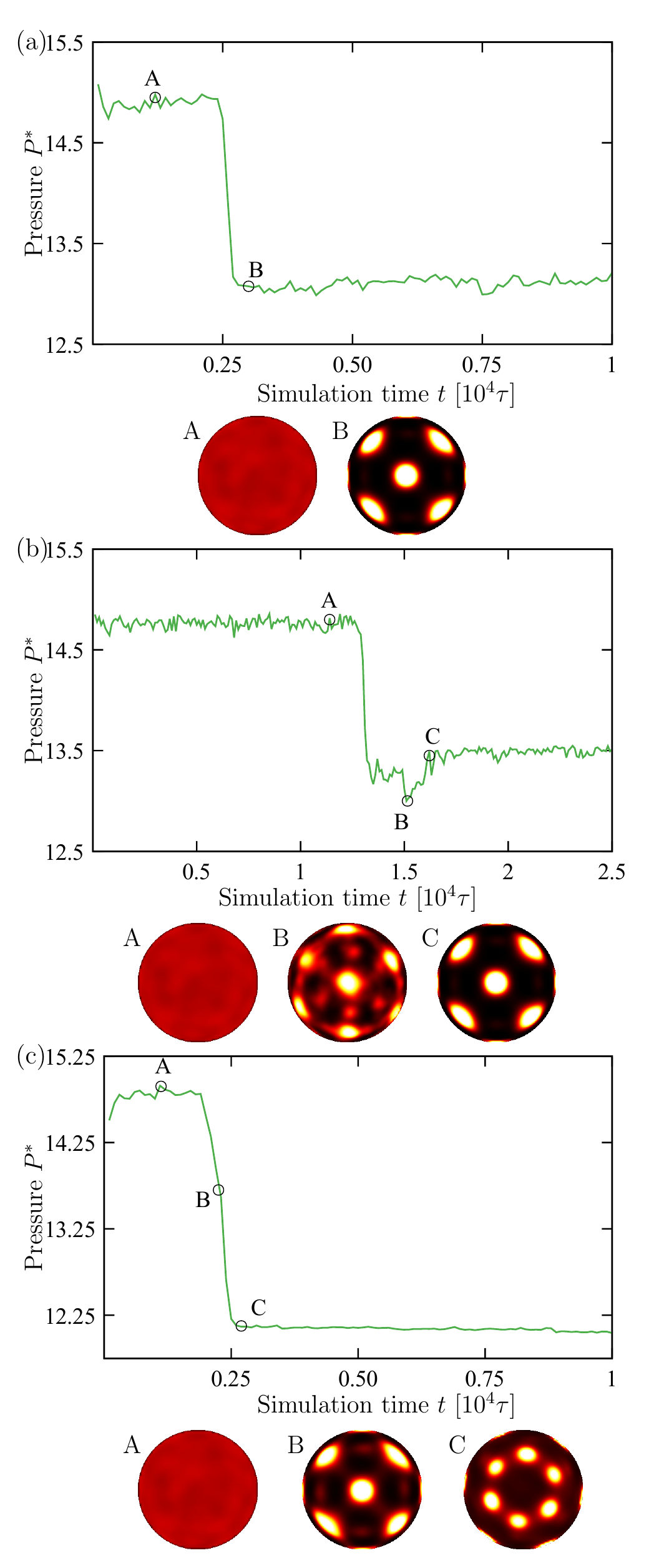}
\caption{Evolution of reduced pressure $P^\ast$ during crystallization at (a)~$(\phi,D)=(0.59,8\%)$, (b)~$(\phi,D)=(0.59,9\%)$, (c)~$(\phi,D)=(0.60,9\%)$.
We observe transitions from the fluid at 'A' (a)~to bcc at 'B', (b)~via $\gamma$-brass at 'B' to bcc at 'C', and (c)~to bcc ('B') and then to fcc ('C').}
\label{fig:Transformation}
\end{figure}

The unit cell is tiled by two elementary building blocks.
Type 1 building blocks are \emph{Laves rhombohedra}.
A Laves rhombohedron is the combination of two Friauf polyhedra (centered at the green particles) and terminated by two tetrahedra at the rhombohedron tips (\cref{fig:CrystalID}(e)).
Laves rhombohedra tile the C14 and C15 Laves phases.
Type 2 building blocks are \emph{decagonal columns}.
A    decagonal column consists of three cylindrical shells: a stacking of joint icosahedra (shell 1; red color), a stacking of interpenetrating Mackay clusters (shell 2; brown color), and an outer cylinder with triangle and hexagon surface features (shell 3)  (\cref{fig:CrystalID}(f)).
The surface features directly attach to the Laves rhombohedra.

%%%%%%%%%%%%%%%%%%%%%%%%%%%%%%%%%%%%%%%%%%%%%%%%%%%%%%%%%%%%%%%%%%%%%%%%%%%%%%%%%%%%%%%%%%%%%%%%%%%%%%%%%%%%%%%%%%%%

\section{Appendix C: Transformation Pathways}

Metastable solid-solid phase transformations are frequently observed in hard sphere mixtures with intermediate size dispersity during crystallization.
Frequently occurring metastable states in simulations are bcc and $\gamma$-brass.
Transformation pathways involving metastable states are typically observed in the range $\phi=(0.58-0.60)$ and $D=(8\%-10\%)$. 
In some of the simulations, bcc crystals remain present even after long simulation times demonstrating a comparably high free energy barrier for a phase transformation from bcc to one of the phases reported in Figs.~1 and~2.
Nevertheless, bcc is only metastable throughout the parameter range studied in this work and will eventually disappear.
In contrast, $\gamma$-brass, always transforms to bcc rapidly indicating a low free energy barrier for this transformation.

Three additional examples of transformation pathways involving metastable crystal phases are shown in \cref{fig:Transformation}.
As indicated by the sharp pressure changes in this figure, solid-solid transformations are usually fast, much faster than the time between successive transformations.
This means the system will usually transform in its entirety before a new transformation starts.
We observe similar transformation pathways independent of system size.

%%%%%%%%%%%%%%%%%%%%%%%%%%%%%%%%%%%%%%%%%%%%%%%%%%%%%%%%%%%%%%%%%%%%%%%%%%%%%%%%%%%%%%%%%%%%%%%%%%%%%%%%%%%%%%%%%%%%

\begin{figure}
\includegraphics[width=0.84\linewidth]{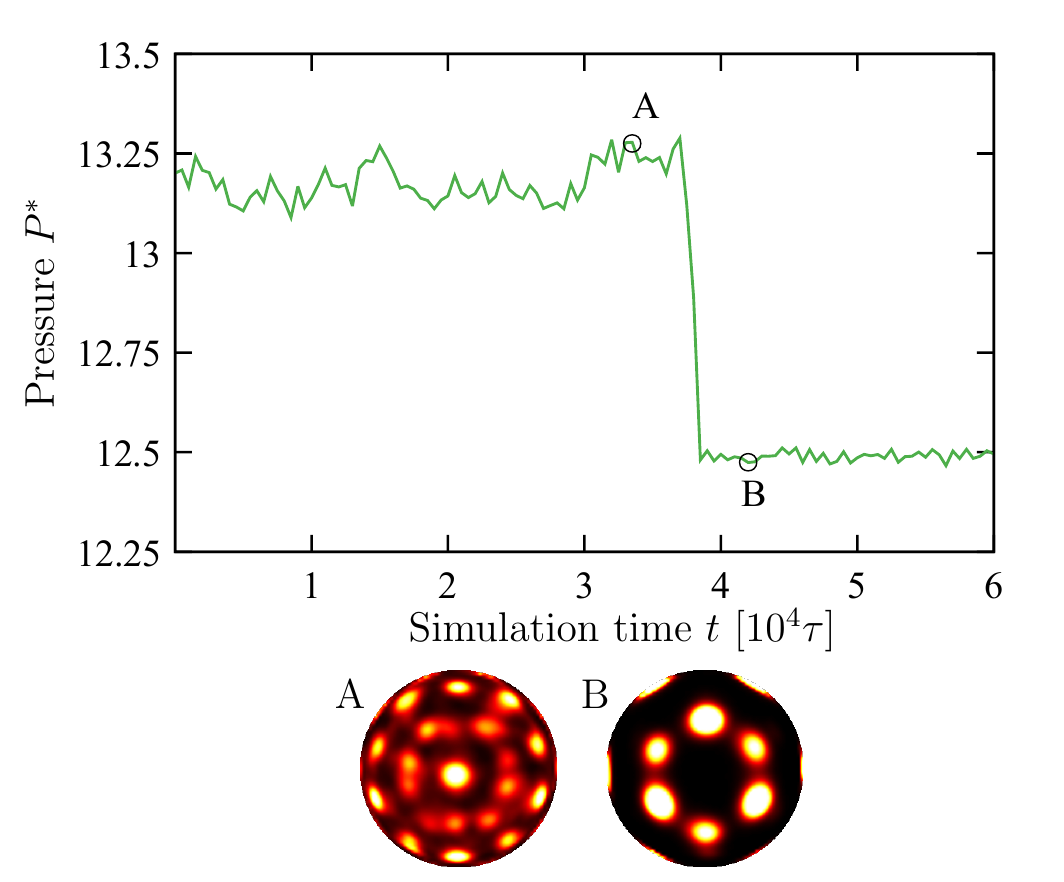}
\caption{Evolution of reduced pressure $P^\ast$ during phase transformation upon lowering packing fraction.
A Frank-Kasper (FK) phase crystallized at parameter pair $(\phi,D)=(0.62,7\%)$ was instantaneously expanded to $(\phi,D)=(0.59,7\%)$.
The equilibrium phase at the new, lower packing fraction is fcc.
Indeed, we observe the transformation from the FK phase at 'A' to fcc at 'B'.
Simulations are performed with particle resize moves.}
\label{fig:FK-fcc}
\end{figure}

\section{Appendix D: Relative thermodynamic stability}

To test relative stability of phases observed in \cref{fig:Resize}(a), we bring a system equilibrated at one parameter pair to another parameter pair where a different crystal structure is predicted to be stable.
In a first step, a configuration at $(\phi,D)=(0.62,7\%)$ is equilibrated using particle resize moves.
The stable phase at this parameter pair is the Frank-Kasper (FK) phase.
In a second step, we convert the configuration instantaneously to $\phi=0.59$ by expanding the simulation box.
At this parameter pair fcc is expected to be stable.
Indeed, our run exhibits a transformation from FK to fcc (\cref{fig:FK-fcc}) demonstrating that fcc is more stable than FK.
The transformation is reproducible in four out of four simulations.

%%%%%%%%%%%%%%%%%%%%%%%%%%%%%%%%%%%%%%%%%%%%%%%%%%%%%%%%%%%%%%%%%%%%%%%%%%%%%%%%%%%%%%%%%%%%%%%%%%%%%%%%%%%%%%%%%%%%

\section{Appendix E: Cloud and shadow curves}

It is helpful to relate phase coexistence, bimodal and spinodal curves of disperse hard sphere systems to the concepts of shadow curves and cloud curves, which were originally developed for the study of multicomponent polymer solutions.
Shadow curves and cloud curves are usually drawn in temperature vs.\ volume fraction plots.
As the temperature of a mixture is decreased, the cloud curve connects temperatures at which the mixture first starts to phase separate.
The shadow curve connects the density of the newly appearing phase at each cloud point temperature.

In hard particle system we can replace temperature with dispersity and volume fraction with packing fraction.
In the language of Ref.~\citenum{Sollich2002}, ``shadow and cloud curves then effectively reduce to the conventional coexistence curves for monodisperse systems''.
This means the cloud curve is identical to the bimodal curve for hard sphere mixtures.
While the shadow curve has meaning in the case of fractionation into multiple fcc crystals, we cannot extend the concept to our complex crystals.
The reason is that the distribution of sphere sizes can have multiple maxima and thus it is not well characterized by a single dispersity value.
The spinodal curve connects points where the system becomes unstable against local small fluctuations.
It is distinct from both the cloud curve and the shadow curve.

%\bibliography{polydisperse}


\end{document}